\documentclass[english,15pt]{article}
\usepackage[T1]{fontenc}
\usepackage[utf8]{inputenc}
\usepackage{geometry}
\usepackage{color}
\usepackage{amsmath}
\usepackage{amssymb}
\usepackage{esint}

\makeatletter
\usepackage{babel}

\usepackage{babel}

\usepackage{babel}

\usepackage{babel}

\makeatother

\usepackage{babel}
\begin{document}

\title{New massless and massive infinite derivative gravity in three dimensions}

\author{Anupam Mazumdar and Georg Stettinger\\
 \\
 Van Swinderen Institute, University of Groningen, Nijenborgh 4, 9747
AG,\\
 Groningen, The Netherlands}

\maketitle
\abstract{ In this paper we will consider the most general quadratic
curvature action with infinitely many covariant derivatives of massless
gravity in three spacetime dimensions. The action is parity invariant
and torsion-free and contains the same off-shell degrees of freedom
as the Einstein-Hilbert action in general relativity. In the ultraviolet,
with an appropriate choice of the propagator given by the \textit{exponential
of an entire function}, the point-like curvature singularity can be
smoothened to a Gaussian distribution, while in the infrared the theory
reduces to general relativity. We will also show how to embed new
massive gravity in ghost-free infinite derivative gravity in Minkowski
background as one of the infrared limits. Finally, we will provide
the tree-level unitarity conditions for infinite derivative gravity
in presence of a cosmological constant in deSitter and Anti-deSitter
spacetimes in three dimensions by perturbing the geometries. }

\section{Introduction}

Einstein's theory of general relativity (GR) contains classical and
quantum singularities in $3+1$ spacetime dimensions~\cite{Hawking:1973uf,tHooft:1974toh}.
Quadratic curvature gravity in four spacetime dimensions ($4d$) indeed
ameliorates the renormalizability issue, but contains massive spin-2
ghosts~\cite{Stelle:1976gc}. A ghost-free theory of quadratic gravity
in $4d$ that contains infinitely many covariant derivatives has been
constructed in Refs.~\cite{Modesto,Kuzmin:1989sp,Tomboulis,Tomboulis-1,Tseytlin:1995uq,Siegel:2003vt,Biswas:2005qr}
and~\cite{Biswas:2011ar,Biswas:2016etb}. The corresponding
action is the most general one which is parity invariant and torsion
free as long as one is interested in linear perturbations around maximally
symmetric spacetimes~\cite{Biswas:2016etb,Biswas:2016egy}. By construction,
the action contains $3$ analytic form factors $F_{i}\left(\Box\right)=\sum_{n}f_{i,n}\Box^{n}$
with infinitely many covariant derivatives representing the Ricci
scalar, the Ricci tensor and the Riemann/Weyl tensor. It was shown
that the full action of ghost-free analytical infinite derivative
theory of gravity (AIDG) in $4d$ can ameliorate the static Schwarzschild
blackhole singularity \cite{Biswas:2011ar,Frolov:2015bia,Frolov,Frolov:2015usa,Edholm:2016hbt,Buoninfante:2018xiw},~\cite{Koshelev:2018hpt,Buoninfante:2018rlq}
and also produces rotating non-singular metrics~\cite{Buoninfante:2018xif}
at linear level. The gravitational interaction weakens
enough that astrophysical objects with even billions of solar masses
may have no singularities, provided certain conditions are met~\cite{Koshelev:2017bxd,Buoninfante:2018rlq}.
The singularities can also be resolved with extended objects such
as static p-branes~\cite{Boos:2018bxf}. At the quantum level, the
interaction introduces non-locality~\cite{Yukawa:1950eq,Efimov:1967pjn,Efimov:1971wx,Efimov:1972wj,Alebastrov:1973vw,Alebastrov:1973np}
and it has been argued that the theory would be power-counting renormalizable~\cite{Tomboulis,Talaganis:2014ida,Modesto}.
A careful analysis of this issue can be found in \cite{Modesto Rachwal,Modesto Rachwal 2}.

It is now wishful to consider whether we could construct ghost-free
AIDG in $3d$, in particular we are motivated to study the gravitational
action in the UV where the higher derivatives play a significant role
around Minkowski and in (A)dS backgrounds. In $3d$, GR itself has
some interesting properties which can be captured by either a metric
theory of gravity or by a Chern-Simons theory~\cite{Townsend,Witten:1988hc,Witten:2007kt}.
It contains $3$ off-shell degrees of freedom but on-shell they do
not survive, hence the physical graviton does not propagate. Furthermore,
there is an interesting connection between $3d$-gravity in AdS and
conformal field theory (CFT) in the boundary~\cite{Maldacena,Kraus:2006wn}
and in AdS$_{3}$ there exists an intriguing non-trivial blackhole
solution~\cite{BTZ,BTZ1}. All these non-trivial features in $3d$
demand further study on the construction of both massless and massive
ghost-free AIDG up to quadratic in curvature.

The aim of this paper will be to construct the conditions on the gravitational
form factors, which will at least guarantee a linearized ghost-free
propagator around Minkowski and (A)dS backgrounds which has massless
Einstein-Hilbert gravity as an IR limit. Finding the propagator in
(A)dS involves non-trivial computations which we will carry out for
the first time in A(dS) for AIDG in $3d$. We will also show how such
a construction can yield new massive gravity~\cite{Bergshoeff:2009hq,Bergshoeff-1}
around Minkowski background, see also ref.~\cite{Ohta:2011rv} for
further classifications of new massive gravity in the IR. We will
restrict ourselves to classical properties and will not consider quantization
of gravity in any of the backgrounds.

The paper is organized as follows: In section two the full equations
of motion of AIDG in $3d$ are discussed and in section three linearized
gravity around Minkowski background and the ghost-free conditions
for the propagator are discussed. In section four we have shown how
AIDG can resolve point like curvature singularities and in section
five, we will discuss how new massive gravity can be treated as an
IR limit of ghost-free AIDG. In section 6, we briefly discuss maximally
symmetric solutions of this action and in section 7, we consider the
conditions for AIDG to be ghost-free in (A)dS backgrounds~%
\footnote{We will use the following conventions:
\begin{itemize}
\item $\eta_{ab}=\mbox{diag}\left(-1,1,1\right)$
\item $a,b,...$ are abstract indices in 3d, $\mu,\nu,...$ are coordinate
indices in 3d and $i,j,...$ are purely spatial coordinate indices
\item $\left(a_{1},...,a_{n}\right)$ and $\left[a_{1},...,a_{n}\right]$
denote (anti-)symmetrization including a factor $\frac{1}{n!}$
\item $c=G_{N}=\hbar=1$ \end{itemize}
}.


\section{The full equations of motion}

The $3d$ analogue of AIDG can be constructed in a very similar fashion
as in $4d$, see for a detailed derivations in $4d$ from the most
general ansatz of parity invariant and torsionless setup in~\cite{Biswas:2011ar,Biswas:2016etb,Biswas:2016egy}.
The action in $3d$ can be captured by the Ricci scalar and the Ricci
tensor, with two form factors:
\begin{equation}
S_{AIDG}=\int d^{3}x\,\sqrt{-g}\,\,\left[\frac{R}{2}+RF_{1}\left(\square\right)R+R_{ab}F_{2}\left(\square\right)R^{ab}\right]\label{IDG}
\end{equation}
where $\Box=g^{ab}\nabla_{a}\nabla_{b}$ has a mass dimension of $2$.
Therefore, $\Box\equiv\Box/M_{s}^{2}$, where $M_{s}$ is a new scale
of gravity in $3d$, beyond which the infinite derivative part becomes
important, while below $M_{s}$ GR becomes a viable option~%
\footnote{We will suppress writing $M_{s}$ in order not to clutter our formulae,
but while discussing physical situation, we will invoke $M_{s}$ and
then we will take care of mentioning it appropriately.%
}. The two form factors are assumed to be analytic and given by an
infinite power series in $\Box$ ~%
\footnote{The $f_{in}$ have to have the dimension $[mass]^{-2-2n}$ according
to our conventions. Further note that here we will only consider analytic
operators of $\Box$, and not non-analytic operators such as $1/\Box$
\cite{Deser:2007jk,Conroy:2014eja} or $\ln(\Box)$ ~\cite{Barvinsky:1985an,Barvinsky:1993en,Donoghue:2014yha}. %
} :
\begin{equation}
F_{i}(\Box)=\sum_{n}f_{i,n}\Box^{n}\,,
\end{equation}
This is the most general form of a covariant action with terms quadratic
in curvature containing infinitely many derivatives with reduces to
GR in the IR regime. The equations of motion can be derived in a very
similar fashion in the $4d$ case, which was first derived in~\cite{Biswas:2013cha}.
The only major difference is that the Weyl tensor is identically zero
in three dimensions, hence there are only two quadratic curvature
terms in the action. Here we present the $3d$ version of that:
\begin{align}
G_{ab}+4G_{ab}F_{1}\left(\square\right)R+g_{ab}RF_{1}\left(\square\right)R-4\left(\nabla_{a}\nabla_{b}-g_{ab}\square\right)F_{1}R-2\Omega_{1ab}\nonumber \\
+g_{ab}\left(\Omega_{1c}^{c}+\overline{\Omega}_{1}\right)+4R_{a}^{c}F_{2}\left(\square\right)R_{cb}-g_{ab}R_{cd}F_{2}\left(\square\right)R^{cd}-4\nabla_{c}\nabla_{b}F_{2}\left(\square\right)R_{a}^{c}\nonumber \\
+2\square F_{2}\left(\square\right)R_{ab}+2g_{ab}\nabla_{c}\nabla_{d}F_{2}\left(\square\right)R^{cd}-2\Omega_{2ab}+g_{ab}\left(\Omega_{2c}^{c}+\overline{\Omega}_{2}\right)-4\Delta_{2ab} & =\tau_{ab}\label{eq:full eom}
\end{align}
where $\tau_{ab}$ is the energy momentum tensor and $G_{ab}$ is
the Einstein tensor. We have defined the symmetric tensors %
\footnote{The notation $A^{\left(l\right)}$is an abbreviation of $\square^{l}A$
for any tensor $A$.%
}
\begin{equation}
\Omega_{1ab}=\overset{\infty}{\underset{n=1}{\sum}}f_{1n}\overset{n-1}{\underset{l=0}{\sum}}\nabla_{a}R^{\left(l\right)}\nabla_{b}R^{\left(n-l-1\right)},\,\,\,\,\,\,\overline{\Omega}_{1}=\overset{\infty}{\underset{n=1}{\sum}}f_{1n}\overset{n-1}{\underset{l=0}{\sum}}R^{\left(l\right)}R^{\left(n-l\right)}
\end{equation}
\begin{equation}
\Omega_{2ab}=\overset{\infty}{\underset{n=1}{\sum}}f_{2n}\overset{n-1}{\underset{l=0}{\sum}}\nabla_{a}R^{cd\left(l\right)}\nabla_{b}R_{cd}^{\left(n-l-1\right)},\,\,\,\,\,\,\overline{\Omega}_{2}=\overset{\infty}{\underset{n=1}{\sum}}f_{2n}\overset{n-1}{\underset{l=0}{\sum}}R^{cd\left(l\right)}R_{cd}^{\left(n-l\right)}
\end{equation}
\begin{equation}
\Delta_{2ab}=\frac{1}{2}\overset{\infty}{\underset{n=1}{\sum}}f_{2n}\overset{n-1}{\underset{l=0}{\sum}}\nabla_{c}\left(R_{d}^{c\left(l\right)}\nabla_{(a}R_{b)}^{d\left(n-l-1\right)}-\nabla_{(a}R^{cd\left(l\right)}R_{b)d}^{\left(n-l-1\right)}\right).
\end{equation}
Obviously, these equations containing all the double sums are very
hard to solve exactly. Nevertheless, one can see that constant curvature
backgrounds are indeed solutions of this theory, i.e.
\[
R={\rm constant},~~R_{ab}={\rm constant}\times g_{ab},
\]
see section 6. In $4d$ such solutions exist, in fact non-trivial
solutions are conformally flat in asymptotically Minkowski background,
see for details~\cite{Buoninfante:2018rlq}. We can also consider
the linearized equations of motion: If we only keep the terms linear
in curvature we obtain
\begin{align}
G_{ab}-4\left(\nabla_{a}\nabla_{b}-g_{ab}\square\right)F_{1}R-4\nabla_{c}\nabla_{b}F_{2}\left(\square\right)R_{a}^{c}\nonumber \\
+2g_{ab}\nabla_{c}\nabla_{d}F_{2}\left(\square\right)R^{cd}+2\square F_{2}\left(\square\right)R_{ab} & =\tau_{ab}.\label{eq:lin full eom}
\end{align}


\section{Unitarity and propagator around Minkowski background}

The linearized limit is particular useful to examine the mechanical
properties of the theory, such as tree-level unitarity and the propagator.
As usual, we will write the metric as %
\begin{equation}
g_{ab}=\eta_{ab}+h_{ab}
\end{equation}
and treat $h_{ab}\ll1$ as a small quantity. Since we want to reduce
the equations of motion to linear order in $h_{ab}$, the action should
contain only terms up to quadratic order in $h_{ab}$ and, moreover,
we expect it to be constructed solely out of $h_{ab}$, $\eta_{ab}$
(the Minkowski metric) and $\partial_{a}$. The most general action
of this kind consists of several terms according to the various index
contractions and reads
\begin{eqnarray}
S_{qua} & = & \frac{1}{4}\int d^{3}x\,\sqrt{-g}\,\,\mbox{\ensuremath{\Biggr[}}\frac{1}{2}h_{ab}\square a\left(\square\right)h^{ab}+h_{b}^{c}b\left(\square\right)\partial_{c}\partial_{a}h^{ab}+hc\left(\square\right)\partial_{a}\partial_{b}h^{ab}\nonumber \\
 &  & +\frac{1}{2}h\square d\left(\square\right)h+h^{cd}\frac{f\left(\square\right)}{2\square}\partial_{c}\partial_{d}\partial_{a}\partial_{b}h^{ab}\mbox{\ensuremath{\Biggl]}}\label{eq:S_lin}
\end{eqnarray}
with analytic functions $a\left(\square\right),...,f\left(\square\right)$
(the exact definitions are merely a convention, of course; extra factors
of $\frac{1}{2}$ or $\square$ are inserted for later convenience).
The resulting linearized equations of motion read
\begin{align}
\frac{1}{2}\square a\left(\square\right)h_{ab}+b\left(\square\right)\partial_{c}\partial_{(a}h_{b)}^{c}+\frac{1}{2}c\left(\square\right)\left(\eta_{ab}\partial_{c}\partial_{d}h^{cd}+\partial_{a}\partial_{b}h\right)\nonumber \\
+\frac{1}{2}\square d\left(\square\right)h\eta_{ab}+\frac{f\left(\square\right)}{2\square}\partial_{a}\partial_{b}\partial_{c}\partial_{d}h^{cd} & =-\tau_{ab}\,.
\end{align}
\footnote{The definition of $\tau_{ab}$ here defers from the previous section
by an unimportant numerical factor.%
} To compute $a\left(\square\right),...,f\left(\square\right)$ in
terms of $F_{1}\left(\square\right)$ and $F_{2}\left(\square\right)$,
we insert the linearized expressions for the curvature quantities
\begin{align}
R_{ab}^{(1)} & =\partial_{c}\partial_{(a}h_{b)}^{c}-\frac{1}{2}\partial_{a}\partial_{b}h-\frac{1}{2}\square h_{ab}\,,\label{eq:lin Ricci}\\
R^{(1)} & =\partial_{a}\partial_{b}h^{ab}-\square h\,,
\end{align}
into Eq.(\ref{eq:lin full eom}) and compare the coefficients of the
different terms. We get~\cite{Biswas:2011ar}:
\begin{align}
a\left(\square\right)= & 1+2F_{2}\left(\square\right)\square=-b\left(\square\right)\,,\nonumber \\
c\left(\square\right)= & 1-8F_{1}\left(\square\right)\square-2F_{2}\left(\square\right)\square=-d\left(\square\right)\,,\nonumber \\
f\left(\square\right)= & 8F_{1}\left(\square\right)\square+4F_{2}\left(\square\right)\square\,.\label{eq:abcdf relations}
\end{align}
The constant terms correspond to the Einstein-Hilbert contribution,
so for $F_{1},\,\, F_{2}\rightarrow0$ we recover pure Einstein gravity.

If we wish to demand that the IR limit of the action Eq.~(\ref{IDG})
is that of Einstein's GR, then similar to the argument provided in
Refs.~\cite{Biswas:2011ar,Biswas:2016etb,Biswas:2016egy}, we want
the equations of motion and hence the propagator to be proportional
to the GR-case, so we demand that $f\left(\square\right)$ should
be zero. As a result, $a\left(\square\right)=c\left(\square\right)$,
and the equations of motion can now be written in momentum space,
using
\begin{equation}
h_{ab}\left(x\right)=\int d^{3}k\, e^{ik_{\nu}x^{\nu}}h_{ab}\left(k\right)\,,
\end{equation}
and
\begin{equation}
\frac{1}{2}a\left(-k^{2}\right)\left(-k^{2}h_{ab}+2k_{c}k_{(a}h_{b)}^{c}-\eta_{ab}k_{c}k_{d}h^{cd}-k_{a}k_{b}h+k^{2}h\eta_{ab}\right)=-\tau_{ab}\,.\label{eq: lin eom in mom space}
\end{equation}
To obtain the free propagator, we have to invert the field equations
which is not possible directly, because they contain zero modes corresponding
to gauge degrees of freedom. An easy way to get rid of the gauge modes
is to use spin projection operators~\cite{Biswas:2011ar,Biswas:2013kla,VanNieuwenhuizen:1973fi},
see Appendix 9.1 for the details. We arrive at the propagator where
the momentum dependent part is given by:
\begin{equation}
\Pi_{AIDG}=\frac{P_{s}^{2}}{a\left(-k^{2}\right)k^{2}}-\frac{P_{s}^{0}}{a\left(-k^{2}\right)k^{2}}=\frac{1}{a\left(-k^{2}\right)}\Pi_{GR}\,.\label{eq:propagator}
\end{equation}
As promised, the propagator is proportional to the GR-propagator and
by choosing $a\left(\square\right)$ in a clever way, namely as exponential
of an \textit{entire function}, we will not introduce any new pole
in the graviton propagator in flat background %
\footnote{There actually exist exceptions to this principle,
e. g. by using complex conjugate poles, see \cite{Modesto 2,Anselmi Piva,Anselmi,Luca et al}%
}. Therefore, by going from UV to IR \textit{only} the 3 dynamical
degrees of freedom, namely the spin-2 and spin-0 components propagate
in a sandwiched propagator, sandwiched between two conserved currents.
Otherwise the propagator would have additional poles associated to
additional particle excitations. The simplest choice is~\cite{Biswas:2011ar}
\begin{equation}
a\left(\square\right)=e^{{-\square}/{M_{s}^{2}}}\,,\label{eq:a box flat}
\end{equation}
with a certain mass scale $M_{s}$ that can be interpreted as the
\emph{scale of non-locality.} The choice of sign in $4d$ was obtained
by demanding that the Newtonian gravitational potential recovers $1/r$
behavior in the IR, see for details~\cite{Talaganis:2014ida}. The
negative sign in the exponent also helps the UV properties which we
will exhibit below. Since the propagator is suppressed in the high
energy regime, there is an indication that the theory may become asymptotically
free. It implies from $a(\Box)=c(\Box)$~\cite{Biswas:2011ar} that
the form factors are now constrained in the Minkowski background:
\begin{eqnarray}
2F_{1}(\Box)+F_{2}(\Box)=0\,,\label{eq:constraint}\\
F_{1}\left(\square\right)=-\frac{e^{{-\square}/{M_{s}^{2}}}-1}{4\square},\,\,\,\,\,\,\quad\quad F_{2}\left(\square\right)=\frac{e^{{-\square}/{M_{s}^{2}}}-1}{2\square}\,.\label{eq:F_in flat space}
\end{eqnarray}
Note that in the low-energy-limit $M_{s}\rightarrow\infty$ the $F_{i}\left(\square\right)$
tend to zero so we get Einstein gravity as expected. There is one
more important issue: The second term in Eq.~(\ref{eq:propagator})
has the wrong sign and therefore indicates the presence of a ghost
state. However, this is an example of a \emph{benign ghost } which
does not spoil unitarity of the associated quantum theory%
\footnote{Since we do not quantize the theory in a rigorous
way, what is shown here is just \emph{tree-level
}unitarity. For a discussion of perturbative unitarity
to all orders see \cite{Briscese Modesto,Chin Tomboulis}%
}. Benign ghosts are a common feature of gauge theories in general.

It is a well known fact that Einstein gravity in $3d$ has no on-shell
propagating degrees of freedom, and since per construction we did
not change the number of local excitations, we expect that statement
still to be true (the derivation can be found in Appendix 9.2). One
should however keep in mind that this is only true on-shell; if we
do not demand the vacuum field equations to hold we can only remove
three degrees of freedom through Eq.~(\ref{eq:gauge trafo}). The
remaining three propagate off-shell as it can be seen in the propagator,
see Eq.~(\ref{eq:propagator}).

Moreover, due to non-locality we conclude that also
causality must be violated in the UV regime. In \cite{Briscese Modesto 2}
it was argued that argued that those violations can never be detected
in any laboratory experiment. We will not go into detail regarding
this issue and refer to the discussions in \cite{Briscese Modesto 2,Tomboulis:2015gfa}.


\section{Adding a Dirac-delta source}

In this section, we wish to show that by adding sources (or more precisely,
a point source) will change the behavior of the solutions drastically.
Let us briefly recall the situation in Einstein gravity: Due to the
local field equations $R_{ab}=0$ space is always flat everywhere,
adding a point source $\tau_{ab}\sim\delta(x^{i})$ (where $x^{i}$
denote the two spatial coordinates) will merely change the behavior
of $R_{ab}$ at $x=0$, leading to a \emph{conical singularity}.

To analyze the problem in AIDG, we wish to work again with the linearized
field equations for $h_{ab}.$ In momentum space, we have %
\begin{equation}
\tau_{ab}\left(k\right)=\int d^{3}x\, e^{-ikx}\delta^{2}\left(x^{i}\right)m\delta_{a}^{0}\delta_{b}^{0}=2\pi m\delta_{a}^{0}\delta_{b}^{0}\delta\left(k^{0}\right)\,,
\end{equation}
Acting on it with the propagator yields
\begin{equation}
\Pi_{ab}^{\,\,\,\,\,\, cd}\tau_{cd}\left(k\right)=h_{ab}\left(k\right)=2\pi m\frac{1}{k^{2}a\left(-k^{2}\right)}\left(\delta_{a}^{0}\delta_{b}^{0}+\eta_{ab}\right)\delta\left(k^{0}\right)\,,
\end{equation}
so $h_{ab}$ takes the form takes the simple form $\begin{pmatrix}0 & 0 & 0\\
0 & \psi & 0\\
0 & 0 & \psi
\end{pmatrix}$ with
\begin{equation}
\psi=\int\frac{d^{3}k}{\left(2\pi\right)^{3}}e^{ikx}\,\delta\left(k^{0}\right)\frac{2\pi m}{k^{2}a\left(-k^{2}\right)}=\int\frac{d^{2}k}{\left(2\pi\right)^{2}}e^{ik_{i}x^{i}}\frac{m}{k^{i}k_{i}a\left(-k^{i}k_{i}\right)}.
\end{equation}
Plugging that into the linearized Ricci tensor Eq.~(\ref{eq:lin Ricci})
yields
\begin{equation}
R_{ab}=-\frac{1}{2}\begin{pmatrix}0 & 0 & 0\\
0 & \Delta\psi & 0\\
0 & 0 & \Delta\psi
\end{pmatrix}\,,
\end{equation}
where $\Delta$ denotes the two dimensional (purely spatial) Laplacian.
$\Delta\psi$ can now be evaluated straight forwardly for our preferred
choice $a=e^{{k^{2}}/{M_{s}^{2}}}$:
\begin{equation}
\Delta\psi=\partial_{i}\partial^{i}\int\frac{d^{2}k}{\left(2\pi\right)^{2}}e^{ik_{i}x^{i}}\frac{m}{k^{i}k_{i}}e^{\frac{-k_{i}k^{i}}{M_{s}^{2}}}=-m\int\frac{d^{2}k}{\left(2\pi\right)^{2}}e^{ik_{i}x^{i}}e^{\frac{-k^{i}k_{i}}{M_{s}^{2}}}=-\frac{M_{s}^{2}m}{4\pi}e^{-\frac{M_{s}^{2}}{4}\left(x_{1}^{2}+x_{2}^{2}\right)}\,.
\end{equation}
So the Ricci tensor turns out to be a \emph{Gaussian distribution}
around the point source. If we take the limit $M_{s}\rightarrow\infty$
the Gaussian turns into a delta distribution and we recover the expected
result of pure Einstein gravity. The infinitely many derivatives have
the effect of \emph{smearing out} the conical singularity and the
Ricci scalar stays finite, in strong analogy with the $4d$ case \cite{Biswas:2011ar}.


\section{New Massive Gravity as an IR limit of AIDG}

So far we have analyzed AIDG under the condition that it reduces to
Einstein gravity in the limit $M_{s}\rightarrow\infty$. In $3d$,
however, there exists another possible low-energy-limit called \emph{New
Massive Gravity, } see \cite{Bergshoeff:2009hq,Bergshoeff-1,Ohta:2011rv}~%
\footnote{ A similar embedding of massive gravity into infinite derivative gravity
has been done in four dimensions, see \cite{Modesto Tsujikawa}.%
}. It consists more or less in Stelle's fourth order theory adapted
to three dimensions, with the action
\begin{equation}
S_{NMG}=\int d^{3}x\,\sqrt{-g}\,\,\left[\frac{R}{2}+\alpha R^{2}+\beta R_{ab}R^{ab}\right]\,,
\end{equation}
(Note that $R$ and $R_{ab}$ here denote the \emph{actual }Ricci
tensor and scalar again, not the background quantities.) In contrast
to four dimensions, there is a possibility to get rid of the Weyl
ghost; namely for the choice $\alpha=-\frac{3}{8}\beta.$ More specifically,
we will consider the action~\cite{Bergshoeff:2009hq,Bergshoeff-1}
\begin{equation}
S_{NMG}=\int d^{3}x\,\sqrt{-g}\,\,\left[-\frac{R}{2}-\frac{3}{8m^{2}}R^{2}+\frac{1}{m^{2}}R_{ab}R^{ab}\right]\,,
\end{equation}
where $m$ is a new mass parameter. Notice that we changed the sign
of the Einstein-Hilbert term deliberately. The propagator can be straight
forwardly evaluated with the help of spin projection operators, and
reads
\begin{equation}
\Pi_{NMG}=-\frac{P_{s}^{2}}{\left(1+\frac{2k^{2}}{m^{2}}\right)k^{2}}+\frac{P_{s}^{0}}{k^{2}}=-\Pi_{GR}+\frac{P_{s}^{2}}{k^{2}+\frac{m^{2}}{2}}\,,
\end{equation}
so we have one additional propagating mode compared to GR which is
a spin two tensor with mass squared $\frac{m^{2}}{2}$. This is the
usual Weyl-ghost familiar from Stelle's theory, however, we have reversed
its sign here so that is has positive energy. As a result, the GR-part
comes with the wrong sign. But this is not a problem since the GR-excitations
do not propagate and the theory is still unitary.

The other open issue is renormalizability: As Stelle has proved, fourth-order
gravity is renormalizable and in three dimensions we still expect
that statement to hold. While this is true in principle, there are
specific combinations of $\alpha$ and $\beta$ which destroy the
renormalizability, namely exactly those which provide unitarity! Hence,
like in four dimensions, we cannot have unitarity and renormalizability
at the same time in fourth-order gravity.

Here our aim is to embed NMG in AIDG, and see how it arises in the
IR. We now want to construct a theory containing infinitely many derivatives
which reduces to NMG in the limit $M_{s}\rightarrow\infty$ and does
not change the particle content. Note that we have now two mass parameters
in the theory and will assume the hierarchy $m\ll M_{s}$. As before,
we want the propagator be proportional to $\Pi_{NMG}$, but suppressed
in the UV-limit. The factor of proportionality must not have any zeros
and shall therefore be of the form $Ce^{\gamma\left(-k^{2}\right)}$
with $C$ a constant and $\gamma\left(-k^{2}\right)$ an entire function.
The AIDG-action will be again of the form
\begin{equation}
S_{AIDG}=\int d^{3}x\,\sqrt{-g}\,\,\left[-\frac{R}{2}+RF_{1}\left(\square\right)R+R_{ab}F_{2}\left(\square\right)R^{ab}\right]\,,
\end{equation}
(with the reversed sign in front of Einstein-Hilbert term again),
and we demand
\begin{equation}
F_{1}\left(\square\right)\rightarrow-\frac{3}{8m^{2}}\,\,\,\,\,\,\,\,\mbox{and}\,\,\,\,\,\,\,\, F_{1}\left(\square\right)\rightarrow\frac{1}{m^{2}}\,,\label{eq:fi0 limits}
\end{equation}
in the limit $M_{s}\rightarrow\infty$. The relations \ref{eq:abcdf relations}
containing the new sign now, read
\begin{align}
a\left(\square\right) & =-1+2F_{2}\left(\square\right)\square=-b\left(\square\right)\,,\nonumber \\
c\left(\square\right) & =-1-8F_{1}\left(\square\right)\square-2F_{2}\left(\square\right)\square=-d\left(\square\right)\,,\nonumber \\
f\left(\square\right) & =8F_{1}\left(\square\right)\square+4F_{2}\left(\square\right)\square\,,
\end{align}
and the propagator (now written in the momentum space) is still given
by
\begin{equation}
\Pi_{AIDG}=\frac{P_{s}^{2}}{a\left(-k^{2}\right)k^{2}}+\frac{P_{s}^{0}}{\left(a\left(-k^{2}\right)-2c\left(-k^{2}\right)\right)k^{2}}=\frac{1}{Ce^{\gamma\left(-k^{2}\right)}}\Pi_{NMG}\,.
\end{equation}
We can now read off the relations
\begin{equation}
a\left(-k^{2}\right)=C\left(1+\frac{2k^{2}}{m^{2}}\right)e^{\gamma\left(-k^{2}\right)}\,\,\,\,\,\,\,\mbox{and}\,\,\,\,\,\,\, a\left(-k^{2}\right)-2c\left(-k^{2}\right)=Ce^{\gamma\left(-k^{2}\right)}\,,
\end{equation}
which implies
\begin{equation}
F_{1}\left(-k^{2}\right)=\frac{Ce^{\gamma\left(-k^{2}\right)}+1}{4k^{2}}+\frac{3}{8}\frac{Ce^{\gamma\left(-k^{2}\right)}}{m^{2}},\,\,\,\,\,\,\,\,\,\, F_{2}\left(-k^{2}\right)=-\frac{Ce^{\gamma\left(-k^{2}\right)}+1}{2k^{2}}-\frac{Ce^{\gamma\left(-k^{2}\right)}}{m^{2}}.
\end{equation}
We see that analyticity of the $F_{i}$ requires $C=-1$. For the
function $\gamma\left(-k^{2}\right)$, we can choose the simplest
analytic possibility $\gamma={k^{2}}/{M_{s}^{2}}$, with $M_{s}$
the scale of non-locality. The form factors then take the form
\begin{equation}
F_{1}\left(-k^{2}\right)=-\frac{e^{\frac{k^{2}}{M_{s}^{2}}}-1}{4k^{2}}-\frac{3}{8}\frac{e^{\frac{k^{2}}{M_{s}^{2}}}}{m^{2}},\,\,\,\,\,\,\,\,\,\, F_{2}\left(-k^{2}\right)=\frac{e^{\frac{k^{2}}{M_{S}^{2}}}-1}{2k^{2}}+\frac{e^{\frac{k^{2}}{M_{S}^{2}}}}{m^{2}}\,,
\end{equation}
and we see that the constant terms are given by
\begin{equation}
f_{10}=-\frac{1}{4M_{s}^{2}}-\frac{3}{8m^{2}},\,\,\,\,\,\,\,\,\,\,\, f_{20}=\frac{1}{2M_{s}^{2}}+\frac{1}{m^{2}}\,,
\end{equation}
and fulfill Eq.(\ref{eq:fi0 limits}) in the limit when $M_{s}\rightarrow\infty$.

Hence, we have constructed a viable infinite derivative extension
of New Massive Gravity which does not alter the particle content.
As per construction it is tree-level unitary, renormalizability has
to be checked separately and no proof is available yet.


\section{Maximally symmetric solutions}

In this section we want to go beyond the linearized limit and study
\emph{maximally symmetric solutions} of the full field equations.
The solutions we find will be important in the consequent chapters
about AIDG in (A)dS-background. In a maximally symmetric spacetime,
the relations
\begin{equation}
R_{abcd}=\frac{R}{6}\left(g_{ac}g_{bd}-g_{ad}g_{bc}\right)\,\,\,\,\,\mbox{and}\,\,\,\,\, R_{ab}=\frac{R}{3}g_{ab}\,,
\end{equation}
hold with $R$ constant over the manifold. That implies that every
curvature quantity is annihilated by the covariant derivative, so
most of the terms in Eq.~(\ref{eq:full eom}) drop out. From the
infinite derivative terms only the zeroth order terms without boxes
denoted as $f_{10}$ and $f_{20}$, contribute. Plugging in the expressions
above yields
\begin{equation}
R^{2}\left(\frac{1}{3}f_{10}+\frac{1}{9}f_{20}\right)-\frac{R}{6}+\Lambda=0\,,\label{eq:qadr R-lambda eq in 3d}
\end{equation}
which is a quadratic equation, so we will in general have \emph{two
}solutions of curvature for a given $\Lambda$. A particularly interesting
case is $\Lambda=0$: Here we get additionally to $R=0$ also the
second solution
\begin{equation}
R=\frac{1}{2f_{10}+\frac{2}{3}f_{20}}\,,
\end{equation}
so an ``effective cosmological constant'' is generated by the higher-order
terms.

One could ask now: Is it then even necessary to include $\Lambda$
in the action? The answer is yes, because, as we will see in section
7.4, the form factors and hence also their zero'th coefficients will
be constrained to certain values%
\footnote{However, the constraints are not the same as derived in sections 3
and 5 because those results were obtained in Minkowski background
with $R=0$.%
}. So if $\Lambda=0$, we have only one specific numerical value available
for the background curvature. To obtain the full variaty of backgrounds,
we have to include $\Lambda.$

The situation is fundamentally different in four dimensions because
the Weyl tensor $C_{abcd}$ is not necessarily zero, so we have to
include a term $C_{abcd}F_{3}\left(\square\right)C^{abcd}$ in the
action. However, $C_{abcd}$ is zero for maximally symmetric spacetimes,
so the additional term does not change the calculation. Computing
Eq.~(\ref{eq:qadr R-lambda eq in 3d}) in a general number of dimensions
$d$ using arbitrary $f_{i0}$ gives
\begin{equation}
R^{2}\left(f_{10}\frac{4-d}{d}+f_{20}\frac{4-d}{d^{2}}\right)+R\frac{2-d}{2d}+\Lambda=0\,,\label{eq:R-lambda, general d}
\end{equation}
and one sees that for $d=4$ the quadratic part completely drops out.
That means in four dimensions we have always the familiar relation
\begin{equation}
R=4\Lambda\,.
\end{equation}
and the background geometry is independent of the form factors $F_{i}\left(\square\right).$
This apparent simplification is absent in three dimensions which leads
to some caveats as can be seen in the next chapter.

As a side remark, we see from these results that also the famous BTZ
black hole is an exact solution: As shown in detail in Ref.~\cite{BTZ,BTZ1},
the BTZ is just an orbifold of AdS space and hence locally indistinguishable
from global AdS. The non-trivial boundary conditions do not cause
any problems and the BTZ is a perfectly viable background for AIDG
in $3d$. We should point out here that the infinitely many derivatives
did not play any role in this section, only the zeroth order terms
entered the calculation. The result Eq.~(\ref{eq:qadr R-lambda eq in 3d})
is also true in Stelle's fourth order gravity.


\section{AIDG in (A)dS(3)}

\subsection{The perturbations around (A)dS}

We would like to derive the linearized equations of motion in a stable
(A)dS background%
\footnote{Recently this analysis was extended to more general
backgrounds like conformally flat spacetimes \cite{Sravan Shubham}.%
}. The whole procedure is in close analogy with Refs.~\cite{Biswas:2016etb,Biswas:2016egy,Sravan et al}.
The action we will consider here is in $3$ dimension, and it is given
by:
\begin{equation}
S_{AIDG}=\int d^{3}x\,\sqrt{-g}\,\,\left[\frac{R}{2}+RF_{1}\left(\square\right)R+R_{ab}F_{2}\left(\square\right)R^{ab}-\Lambda\right],
\end{equation}
In this paper, we will consider the details of the scalar, vector
and tensor decomposition of the quadratic part of the action around
(A)dS. For later convenience we will rewrite this action using the
\emph{traceless Ricci tensor}
\[
S_{ab}=R_{ab}-\frac{1}{3}g_{ab}R,
\]
as
\begin{equation}
S_{AIDG}=\int d^{3}x\,\sqrt{-g}\,\,\left[\frac{R}{2}+R\widehat{F}_{1}\left(\square\right)R+S_{ab}\widehat{F}_{2}\left(\square\right)S^{ab}-\Lambda\right].\label{IDG-1}
\end{equation}
This amounts just to a trivial redefinition of the $F_{i}$'s, we
obtain:
\begin{equation}
\widehat{F}_{1}\left(\square\right)=F_{1}\left(\square\right)+\frac{1}{3}F_{2}\left(\square\right),\,\,\,\,\,\,\,\,\,\,\,\quad\quad\quad\,\,\,\widehat{F}_{2}\left(\square\right)=F_{2}\left(\square\right).\label{redefine}
\end{equation}
To obtain the linearized equations of motion, we have to compute the
\emph{second variation, }which is a straight forward but laborious
task. We have to replace all the quantities by their second order
perturbation (for the details see Appendix 9.3) using
\begin{equation}
g_{ab}=\overline{g}_{ab}+h_{ab}\,,
\end{equation}
and keep only terms quadratic in $h_{ab}$. The bars on the background
quantities have been omitted for simplicity. The different parts of
the action shall be analyzed separately.


\subsubsection{Einstein-Hilbert part of the action including $\Lambda$}

The pure Einstein-Hilbert part of the action from Eq.~(\ref{IDG-1})
becomes
\begin{equation}
S_{EH}\simeq\int d^{3}x\,\sqrt{-g}\left(1+\frac{h}{2}+\frac{h^{2}}{8}-\frac{h_{ab}h^{ab}}{4}\right)\left[\frac{1}{2}\left(R_{ab}+\delta R_{ab}+\delta^{2}R_{ab}\right)\left(g^{ab}-h^{ab}+h^{ac}h_{c}^{b}\right)-\Lambda\right]\,,
\end{equation}
where $\delta R_{ab}$ and $\delta^{2}R_{ab}$ are the first and second
order variations of the Ricci tensor, defined in equation Eq.~(\ref{eq:Ricci pert}).
After a lengthy calculation outlined in Appendix 9.4, collecting all
the quadratic terms yields
\begin{align}
\delta^{2}S_{EH}= & \int d^{3}x\,\sqrt{-g}\mbox{\ensuremath{\Biggr[}}\frac{1}{8}h^{ab}\square h_{ab}-\frac{1}{8}h\square h+\frac{1}{4}h\nabla_{a}\nabla_{b}h^{ab}+\nonumber \\
 & \frac{1}{4}\nabla_{a}h^{ab}\nabla_{c}h_{b}^{c}+\frac{1}{48}h^{2}R-\frac{1}{12}h_{ab}h^{ab}R-\Lambda\left(\frac{h^{2}}{8}-\frac{h_{ab}h^{ab}}{4}\right)\mbox{\ensuremath{\Biggl]}}.\label{eq:delta squared EH}
\end{align}
For future purposes we shall write this quantity as
\begin{equation}
\delta^{2}S_{EH}\equiv\frac{1}{2}\int d^{3}x\,\sqrt{-g}\delta_{0}\,,
\end{equation}
with
\begin{equation}
\delta_{0}=\delta^{2}R+\frac{h}{2}\delta R+\left(\frac{h^{2}}{8}-\frac{h_{ab}h^{ab}}{4}\right)\left(R-2\Lambda\right).\label{eq:delta_0 Gleichung}
\end{equation}


\subsubsection{Terms containing $\widehat{F}_{1}(\Box)$}

The part of the action containing the Ricci scalar reads:
\begin{align}
S_{R}\simeq & \int d^{3}x\,\sqrt{-g}\left(1+\frac{h}{2}+\frac{h^{2}}{8}-\frac{h_{ab}h^{ab}}{4}\right)\left(R+\delta R+\delta^{2}R\right)\nonumber \\
 & \left(\widehat{F}_{1}(\square)+\delta\widehat{F}_{1}(\square)+\delta^{2}\widehat{F}_{1}(\square)\right)\left(R+\delta R+\delta^{2}R\right).
\end{align}
Again, collecting all terms quadratic in $h_{ab}$ results in
\begin{eqnarray}
\delta^{2}S_{R} & = & \int d^{3}x\,\sqrt{-g}\mbox{\ensuremath{\Biggr[}}R\widehat{F}_{1}\left(\square\right)\delta^{2}R+R\delta\widehat{F}_{1}\left(\square\right)\delta R+R\delta^{2}\widehat{F}_{1}\left(\square\right)R+\widehat{F}_{1}\left(\square\right)\delta R\nonumber \\
 &  & +\delta R\delta\widehat{F}_{1}\left(\square\right)R+\delta^{2}R\widehat{F}_{1}\left(\square\right)R+\frac{h}{2}\left(R\widehat{F}_{1}\left(\square\right)\delta R+R\delta\widehat{F}_{1}\left(\square\right)R+\delta R\widehat{F}_{1}\left(\square\right)R\right)\nonumber \\
 &  & +\left(\frac{h^{2}}{8}-\frac{h_{ab}h^{ab}}{4}\right)R\widehat{F}_{1}\left(\square\right)R\mbox{\ensuremath{\Biggl]}}.
\end{eqnarray}
The variation of $\square$ acting on a scalar is given by
\begin{equation}
\delta\left(\square\right)\varphi=\left(-h^{ab}\nabla_{a}\partial_{b}-g^{ab}\delta\Gamma_{ab}^{c}\partial_{c}\right)\varphi\,,\label{eq:var box}
\end{equation}
where $\delta\Gamma_{ab}^{c}$ denotes the variation of the Christoffel
symbol (see Appendix 9.3). We conclude that the constant background
curvature $R$ is annihilated by all variations, $\delta^{i}\widehat{F}_{1}\left(\square\right)$,
but the zeroth coefficient $\widehat{f}_{10}$ in the expansion of
$\widehat{F}_{1}\left(\square\right)=\sum_{n=0}^{\infty}\widehat{f}_{1n}\square^{n}$
survives. A reorganization of the terms now yields
\begin{eqnarray}
\delta^{2}S_{R} & = & \int d^{3}x\,\sqrt{-g}\mbox{\ensuremath{\Biggr[}}\left(h\delta R+\left(\frac{h^{2}}{8}-\frac{h_{ab}h^{ab}}{4}\right)R+2\delta^{2}R\right)\widehat{f}_{10}R+\delta R\widehat{F}_{1}\left(\square\right)\delta R\nonumber \\
 &  & +\frac{h}{2}R\left(\widehat{F}_{1}\left(\square\right)-\widehat{f}_{10}\right)\delta R+R\delta\widehat{F}_{1}\left(\square\right)\delta R\mbox{\ensuremath{\Biggl]}}.
\end{eqnarray}
It can be further simplified by using Eq.~(\ref{eq:delta_0 Gleichung}),
and we arrive at
\begin{eqnarray}
\delta^{2}S_{R} & = & \int d^{3}x\,\sqrt{-g}\mbox{\ensuremath{\Biggr[}}2\widehat{f}_{10}R\delta_{0}+2\widehat{f}_{10}R\left(2\Lambda-\frac{R}{2}\right)\left(\frac{h^{2}}{8}-\frac{h_{ab}h^{ab}}{4}\right)+\delta R\widehat{F}_{1}\left(\square\right)\delta R\nonumber \\
 &  & +\frac{h}{2}R\left(\widehat{F}_{1}\left(\square\right)-\widehat{f}_{10}\right)\delta R+R\delta\widehat{F}_{1}\left(\square\right)\delta R\mbox{\ensuremath{\Biggl]}}.
\end{eqnarray}
\footnote{In four dimensions, the second term vanishes due to the background
constraint $\Lambda=\frac{R}{4}$. %
} In the second line, the two terms can be seen to cancel away. First,
the variation in the last term has to appear at the extreme left,
otherwise the term becomes a total derivative. Next, by expanding
the power series in both of the terms gives:
\begin{equation}
\sum_{n=1}^{\infty}\widehat{f}_{1n}R\int d^{3}x\,\sqrt{-g}\left(\frac{h}{2}\square+\delta\left(\square\right)\right)\square^{n-1}\delta R\,,
\end{equation}
and by using Eq.~(\ref{eq:var box}) again, we find %
\begin{align}
 & \sum_{n=1}^{\infty}\widehat{f}_{1n}R\int d^{3}x\,\sqrt{-g}\left(\frac{h}{2}\square-h^{ab}\nabla_{a}\partial_{b}-g^{ab}\delta\Gamma_{ab}^{c}\partial_{c}\right)\square^{n-1}\delta R\nonumber \\
= & \sum_{n=1}^{\infty}\widehat{f}_{1n}R\int d^{3}x\,\sqrt{-g}\left(\frac{h}{2}\square-h^{ab}\nabla_{a}\partial_{b}-\nabla_{a}h^{ab}\partial_{b}+\nabla_{a}h\partial^{a}\right)\square^{n-1}\delta R\nonumber \\
= & \sum_{n=1}^{\infty}\widehat{f}_{1n}R\int d^{3}x\,\sqrt{-g}\left(\frac{h}{2}\square^{n}\delta R-\nabla_{a}\left(h^{ab}\partial_{b}\square^{n-1}\delta R\right)-h\nabla_{a}\partial^{a}\square^{n-1}\delta R\right)\nonumber \\
= & \sum_{n=1}^{\infty}\widehat{f}_{1n}R\int d^{3}x\,\sqrt{-g}\left(-\nabla_{a}\left(h^{ab}\partial_{b}\square^{n-1}\delta R\right)\right)\,,
\end{align}
which is a total derivative and therefore vanishes. Hence the final
result is given by:
\begin{equation}
\delta^{2}S_{R}=\int d^{3}x\,\sqrt{-g}\left[2\widehat{f}_{10}R\delta_{0}+2\widehat{f}_{10}R\left(2\Lambda-\frac{R}{2}\right)\left(\frac{h^{2}}{8}-\frac{h_{ab}h^{ab}}{4}\right)+\delta R\widehat{F}_{1}\left(\square\right)\delta R\right].
\end{equation}


\subsubsection{Terms containing $\widehat{F}_{2}(\Box)$ }

The last variation is particularly simple because the traceless Ricci
tensor vanishes for maximally symmetric spacetimes, so the two variations
have to act on both $S_{ab}$ to produce a non-zero result. We have
\begin{align}
\delta^{2}S_{S} & =\int d^{3}x\,\sqrt{-g}\left(\delta R_{ab}-\frac{1}{3}g_{ab}\delta R\right)\widehat{F}_{2}\left(\square\right)\left(\delta R^{ab}-\frac{1}{3}g^{ab}\delta R\right)\nonumber \\
 & =\int d^{3}x\,\sqrt{-g}\left[\delta R_{ab}\widehat{F}_{2}\left(\square\right)\delta R^{ab}-\frac{1}{3}\delta R\widehat{F}_{2}\left(\square\right)\delta R\right]\,,
\end{align}
so that we can write the complete variation as
\begin{eqnarray}
\delta^{2}S & = & \int d^{3}x\,\sqrt{-g}\mbox{\ensuremath{\Biggl[}}\left(\frac{1}{2}+2\widehat{f}_{10}R\right)\delta_{0}+\widehat{f}_{10}R\left(4\Lambda-R\right)\left(\frac{h^{2}}{8}-\frac{h_{ab}h^{ab}}{4}\right)\nonumber \\
 &  & +\delta RF_{1}\left(\square\right)\delta R+\delta R_{ab}F_{2}\left(\square\right)\delta R^{ab}\mbox{\ensuremath{\Biggr]}}.
\end{eqnarray}
using the unhatted $F_{i}\left(\square\right)$ again. This expression
can be simplified considerably by inserting the relation Eq.~(\ref{eq:qadr R-lambda eq in 3d})
between the cosmological constant and the background curvature. Some
of the terms of higher order in $R$ cancel away and we are left with
\begin{equation}
\delta^{2}S=\int d^{3}x\,\sqrt{-g}\left[\left(\frac{1}{2}+\widehat{f}_{10}R\right)\widetilde{\delta}_{0}+\delta RF_{1}\left(\square\right)\delta R+\delta R_{ab}F_{2}\left(\square\right)\delta R^{ab}\right]\,,\label{eq:second variation complete ac}
\end{equation}
where $\widetilde{\delta}_{0}$ is defined as
\begin{equation}
\widetilde{\delta}_{0}=\frac{1}{4}h^{ab}\square h_{ab}-\frac{1}{4}h\square h+\frac{1}{2}h\nabla_{a}\nabla_{b}h^{ab}+\frac{1}{2}\nabla_{a}h^{ab}\nabla_{c}h_{b}^{c}-\frac{R}{12}h_{ab}h^{ab}.\label{eq:delta schlange}
\end{equation}
Note that $\widetilde{\delta}_{0}$ is exactly what we have had obtained
in pure Einstein gravity: Eq. (\ref{eq:qadr R-lambda eq in 3d}) would
then reduce to $\Lambda=6R$ and from Eq. (\ref{eq:delta squared EH})
we would obtain
\begin{equation}
\delta^{2}S_{EH}\equiv\frac{1}{2}\int d^{3}x\,\sqrt{-g}\widetilde{\delta}_{0}.
\end{equation}
So the terms generated by the non-trivial relation Eq. (\ref{eq:qadr R-lambda eq in 3d})
cancel away and Eq. (\ref{eq:second variation complete ac}) has the
same form as in 4d, apart from the fact that the Weyl term is absent.


\subsection{Scalar, vector and tensor decompositions of the metric perturbations }

To proceed further, we have to decompose the metric perturbation into
the different spin states~\cite{Biswas:2016etb,Biswas:2016egy}.
However, there is one big difference: In an AdS background we cannot
go to Fourier space globally, hence the group theoretic arguments
outlined in Appendix 9.1 do not work straight forwardly anymore. Instead,
we will follow the procedure outlined in~\cite{DHoker:1999bve,Biswas:2016etb,Biswas:2016egy},
see also Refs.~\cite{Miao:2011fc,Kahya:2011sy,Allen:1986ta,Antoniadis:1986sb}
where the authors have found the graviton propagator in dS in $4$
dimensions. Let us define the metric perturbation as
\begin{equation}
h_{ab}=h_{ab}^{\bot}+\nabla_{a}A_{b}^{\bot}+\nabla_{b}A_{a}^{\bot}+\nabla_{a}\nabla_{b}B-g_{ab}\phi\,,\label{eq:decompost}
\end{equation}
where the tensor part obeys the transverse and traceless condition
\[
\nabla^{a}h_{ab}^{\bot}=h^{\bot}=0,
\]
and so does the vector part $\nabla^{a}A_{a}^{\bot}=0$. This decomposition
corresponds exactly to the one which was done in flat space using
the spin projection operators, i.e. $h_{ab}^{\bot}$ corresponds to
$P_{s}^{2}$, $A_{a}^{\bot}$ corresponds to $P_{w}^{1}$, $B$ corresponds
to $P_{w}^{0}$ and $\phi$ to $P_{s}^{0}$. Since we do not want
to increase the number of degrees of freedom as compared to Einstein's
gravity, we have to demand that $A_{a}^{\bot}$ and $B$ drop out
of the quadratic action Eq.~(\ref{eq:second variation complete ac}).
This we will show explicitly below. Then, $h_{ab}^{\bot}$ and $\phi$
will correspond exactly to the $3$ off-shell propagating degrees
of freedom. Let us start with decomposing the variation of the Ricci
tensor $\delta R_{ab}$ which appears in the higher-derivative terms.
$\delta R$ can then be obtained by a simple contraction.


\subsubsection{Decomposition of $\delta R_{ab}$}

We note that the content of this subsection is entirely geometrical,
without referring to any particular theory.
\begin{itemize}
\item We start with the vector mode $A_{b}^{\bot}$; inserting Eq. (\ref{eq:decompost})
into Eq.~(\ref{eq: Riemann pert}) (see Appendix 9.3) and contracting
with $\delta_{a}^{c}$ yields
\begin{eqnarray}
\delta R_{\,\,\, b}^{a}\left(A^{\bot}\right) & = & -\frac{R}{12}\left(\nabla_{b}A^{a}+\nabla^{a}A_{b}\right)+\frac{1}{2}\mbox{\ensuremath{\Bigr(}}\nabla_{c}\nabla^{a}\nabla_{b}A^{c}+\nabla_{c}\nabla^{a}\nabla^{c}A_{b}\nonumber \\
 &  & -\square\nabla_{b}A^{a}-\square\nabla^{a}A_{b}+\nabla_{b}\square A^{a}+\nabla_{b}\nabla^{c}\nabla^{a}A_{c}\mbox{\ensuremath{\Bigl)}},
\end{eqnarray}
already using $\nabla_{a}A^{a}=0$. With the help of the Riemann tensor
substitution and the commutation relation Eq.~(\ref{eq:commutation 2})
in Appendix 9.5, we can rewrite this as
\begin{eqnarray}
\delta R_{\,\,\, b}^{a}\left(A^{\bot}\right) & = & -\frac{R}{12}\left(\nabla_{b}A^{a}+\nabla^{a}A_{b}\right)+\frac{1}{2}\Bigr(\nabla^{a}\nabla_{c}\nabla_{b}A^{c}+R_{c\,\,\,\, b}^{\,\,\, a\,\,\, d}\nabla_{d}A^{c}\nonumber \\
 &  & +R_{c\,\,\,\,\,\, d}^{\,\,\, ac}\nabla_{b}A^{d}+R_{\,\,\,\,\,\, bd}^{ac}\nabla_{c}A^{d}-\frac{R}{3}\left(\nabla_{b}A^{a}+\nabla^{a}A_{b}\right)+R_{\,\,\,\,\,\, cd}^{ca}\nabla_{b}A^{d}\mbox{\ensuremath{\Bigl)}}\nonumber \\
 & = & -\frac{R}{12}\left(\nabla_{b}A^{a}+\nabla^{a}A_{b}\right)\nonumber \\
 &  & +\frac{1}{2}\left(\frac{R}{3}\nabla^{a}A_{b}+\frac{R}{6}\nabla^{a}A_{b}+\frac{R}{3}\nabla_{b}A^{a}-\frac{R}{6}\nabla_{b}A^{a}-\frac{R}{3}\left(\nabla_{b}A^{a}+\nabla^{a}A_{b}\right)+\frac{R}{3}\nabla_{b}A^{a}\right)\nonumber \\
 & = & 0\,,
\end{eqnarray}
which is zero as desired.
\item For the scalar mode $B$, we again insert Eq. (\ref{eq:decompost})
into Eq. (\ref{eq: Riemann pert}) and contract with $\delta_{a}^{c}$
to get
\begin{equation}
\delta R_{\,\,\, b}^{a}\left(B\right)=-\frac{R}{12}\left(\nabla^{a}\nabla_{b}B+\delta_{b}^{a}\square B\right)+\frac{1}{2}\left(\nabla_{c}\nabla^{a}\nabla^{c}\nabla_{b}B+\square\nabla^{a}\nabla_{b}B-\nabla_{b}\nabla^{a}\square B+\nabla_{b}\nabla^{c}\nabla^{a}\nabla_{c}B\right)\,.
\end{equation}
Exchanging $\nabla^{a}$ and $\nabla^{c}$ in the first and the last
term of the second parenthesis yields %
\begin{eqnarray}
\delta R_{\,\,\, b}^{a}\left(B\right) & = & -\frac{R}{12}\left(\nabla^{a}\nabla_{b}B+\delta_{b}^{a}\square B\right)+\frac{1}{2}\left(R_{\,\,\,\,\, b}^{ac\,\,\, d}\nabla_{c}\nabla_{d}B+R_{\,\,\,\,\, c}^{ca\,\,\, d}\nabla_{b}\nabla_{d}B\right)\nonumber \\
 & = & -\frac{R}{12}\left(\nabla^{a}\nabla_{b}B+\delta_{b}^{a}\square B\right)+\frac{1}{2}\left(\frac{R}{6}\delta_{b}^{a}\square B-\frac{R}{6}\nabla^{a}\nabla_{b}B+\frac{R}{3}\nabla^{a}\nabla_{b}B\right)\nonumber \\
 & = & 0\,.
\end{eqnarray}

\item For the remaining two modes we expect a non-zero result: The scalar
$\phi$ inserted into \ref{eq: Riemann pert} yields
\begin{equation}
\delta R_{\,\,\,\,\, cd}^{ab}\left(\phi\right)=\frac{R}{6}\delta_{cd}^{ab}\phi-\frac{1}{2}\left(\nabla_{c}\nabla^{b}\delta_{d}^{a}-\nabla_{c}\nabla^{a}\delta_{d}^{b}-\nabla_{d}\nabla^{b}\delta_{c}^{a}+\nabla_{d}\nabla^{a}\delta_{c}^{b}\right)\phi.
\end{equation}
It is now practical to define the traceless differential operator
\begin{equation}
D_{b}^{a}=\nabla_{b}\nabla^{a}-\delta_{b}^{a}\frac{\square}{3}.\label{Dab}
\end{equation}
Using $D_{b}^{a}$ the equation above can be rewritten as
\begin{equation}
\delta R_{\,\,\,\,\, cd}^{ab}\left(\phi\right)=\frac{1}{2}\left(D_{c}^{a}\delta_{d}^{b}+D_{d}^{b}\delta_{c}^{a}-D_{c}^{b}\delta_{d}^{a}-D_{d}^{a}\delta_{c}^{b}\right)\phi+\frac{2\square+R}{6}\delta_{cd}^{ab}\phi\,,
\end{equation}
and %
\begin{equation}
\delta R_{\,\,\, b}^{a}\left(\phi\right)=\left(\frac{1}{2}D_{b}^{a}+\frac{R+2\square}{3}\delta_{b}^{a}\right)\phi\,\,\,\,\,\,\mbox{and}\,\,\,\,\,\,\delta R\left(\phi\right)=\left(2\square+R\right)\phi\,,
\end{equation}
follow straight forwardly.
\item Similarly, the tensor mode gives
\begin{eqnarray}
\delta R_{\,\,\,\,\, cd}^{ab} & = & \frac{R}{12}\left(\delta_{d}^{a}h_{c}^{\bot b}-\delta_{c}^{a}h_{d}^{\bot b}+\delta_{c}^{b}h_{d}^{\bot a}-\delta_{d}^{b}h_{c}^{\bot a}\right)+\nonumber \\
 &  & \frac{1}{2}\left(\nabla_{c}\nabla^{b}h_{d}^{\bot a}-\nabla_{c}\nabla^{a}h_{d}^{\bot b}-\nabla_{d}\nabla^{b}h_{c}^{\bot a}+\nabla_{d}\nabla^{a}h_{c}^{\bot b}\right)\,,
\end{eqnarray}
and after contracting with $\delta_{a}^{c}$, and using the Riemann
tensor substitution
\begin{equation}
\left[\nabla^{c},\nabla_{a}\right]h_{cb}^{\bot}=R_{\,\,\,\,\, ca}^{cd}h_{db}^{\bot}+R_{b\,\,\,\,\, a}^{\,\, dc}h_{cd}^{\bot}\,,
\end{equation}
we obtain
\begin{equation}
\delta R_{\,\,\, b}^{a}\left(h^{\bot}\right)=-\frac{R}{12}h_{b}^{\bot a}-\frac{1}{2}\square h_{b}^{\bot a}+\frac{1}{2}\nabla^{c}\nabla^{a}h_{cb}^{\bot}=-\frac{1}{2}\left(\square-\frac{R}{3}\right)h_{b}^{\bot a}\,,\,\,\,\,\,\,\mbox{and}\,\,\,\,\,\,\delta R\left(h^{\bot}\right)=0\,.\label{eq:deltaRab hab}
\end{equation}

\end{itemize}

\subsubsection{Decomposition of the Einstein-Hilbert part}

Of course we want to find the same non-vanishing degrees of freedom
also in the pure GR-part. Though, it is a well-known result that linearized
Einstein gravity does not contain any longitudinal excitations, we
will verify it here explicitly.
\begin{itemize}
\item By inserting Eq.~(\ref{eq:decompost}) into $\delta^{2}S_{EH}$,
we obtain for the vector mode
\begin{align}
\widetilde{\delta}_{0}\left(A_{a}^{\bot}\right)= & \frac{1}{4}\left(\nabla_{a}A_{b}+\nabla_{b}A_{a}\right)\square\left(\nabla^{a}A^{b}+\nabla^{b}A^{a}\right)+\frac{1}{2}\left(\square A^{b}+\nabla_{a}\nabla^{b}A^{a}\right)\left(\square A_{b}+\nabla_{a}\nabla_{b}A^{a}\right)\nonumber \\
 & -\frac{R}{12}\left(\nabla_{a}A_{b}+\nabla_{b}A_{a}\right)\left(\nabla^{a}A^{b}+\nabla^{b}A^{a}\right).
\end{align}
$\widetilde{\delta}_{0}$ is an integrand, so we can perform partial
integration, moreover utilizing the commutation relations Eqs.~(\ref{eq:commutation 2},
\ref{eq:commutation 3}), we obtain:
\begin{equation}
\widetilde{\delta}_{0}\left(A_{a}^{\bot}\right)=-\frac{1}{2}A_{b}\nabla_{a}\square\left(\nabla^{a}A^{b}+\nabla^{b}A^{a}\right)+\frac{1}{2}A^{b}\left(\square+\frac{R}{3}\right)^{2}A_{b}+\frac{R}{6}A^{b}\left(\square+\frac{R}{3}\right)A_{b}.
\end{equation}
The first term can be further simplified using Eqs.~(\ref{eq:commutation 4},
\ref{eq:commutation 3}):
\begin{equation}
\widetilde{\delta}_{0}\left(A_{a}^{\bot}\right)=-\frac{1}{2}A_{b}\left(\square+\frac{2R}{3}\right)\left(\square+\frac{R}{3}\right)A^{b}+\frac{1}{2}A^{b}\left(\square+\frac{R}{3}\right)^{2}A_{b}+\frac{R}{6}A^{b}\left(\square+\frac{R}{3}\right)A_{b}=0\,.
\end{equation}

\item For the scalar mode $B$, we obtain similarly
\begin{eqnarray}
\widetilde{\delta}_{0}\left(B\right) & = & \frac{1}{4}\nabla_{a}\nabla_{b}B\square\nabla^{a}\nabla^{b}B-\frac{1}{4}\square B\square^{2}B+\frac{1}{2}\square B\nabla_{a}\nabla_{b}\nabla^{a}\nabla^{b}B+\nonumber \\
 &  & \frac{1}{2}\square\nabla_{a}B\square\nabla^{a}B-\frac{1}{12}R\nabla_{a}\nabla_{b}B\nabla^{a}\nabla^{b}B\nonumber \\
 & = & B\mbox{\ensuremath{\Biggl(}}\frac{1}{4}\nabla_{b}\left(\square+\frac{2R}{3}\right)\square\nabla^{b}-\frac{R}{12}\square^{2}+\frac{1}{4}\square^{3}+\nonumber \\
 &  & \frac{1}{2}\square\nabla_{a}R_{\,\,\, b}^{a}\nabla^{b}-\frac{1}{2}\left(\square+\frac{R}{3}\right)\nabla_{a}\square\nabla^{a}-\frac{1}{12}R\left(\square+\frac{R}{3}\right)\square\mbox{\ensuremath{\Biggr)}}B\nonumber \\
 & = & B\left(-\frac{1}{4}\left(\square+\frac{R}{3}\right)^{2}\square+\frac{R}{6}\left(\square+\frac{R}{3}\right)\square+\frac{1}{4}\square^{3}-\frac{1}{36}R^{2}\square\right)B\nonumber \\
 & = & 0\,.
\end{eqnarray}

\item The non-zero modes can also be evaluated straight forwardly resulting
in
\begin{equation}
\widetilde{\delta}_{0}\left(\phi\right)=\frac{3}{4}\phi\square\phi-\frac{9}{4}\phi\square\phi+\frac{3}{2}\phi\square\phi+\frac{1}{2}\nabla_{a}\phi\nabla^{a}\phi-\frac{R}{4}\phi^{2}=-\frac{1}{4}\phi\left(2\square+R\right)\phi\,,
\end{equation}
and
\begin{equation}
\widetilde{\delta}_{0}\left(h^{\bot}\right)=\frac{1}{4}h_{ab}^{\bot}\left(\square-\frac{R}{3}\right)h^{\bot ab}\,.\label{eq:delta0 hab}
\end{equation}

\end{itemize}

\subsection{Propagator of AIDG in (A)dS(3)}

Finally we wish to obtain the propagators for the two remaining modes
$h_{ab}^{\bot}$ and $\phi$. The first question is, if those two
modes really decouple from each other, i.e. that we can write the
final quadratic action as a sum of the two separate actions. We know
this to be true for the GR-part and in the $F_{1}$-term no $h_{ab}^{\bot}$
can survive. Hence, the only suspicious term is the $F_{2}$- term,
but here we can show straight forwardly that no coupling occurs: (see
also Ref.~\cite{Biswas:2016egy}). The question is now: do any terms
survive in the combination
\begin{equation}
\int d^{3}x\,\sqrt{-g}\,\delta R_{ab}\left(\phi\right)F_{2}\left(\square\right)\delta R^{ab}\left(h^{\bot}\right)=-\int d^{3}x\,\sqrt{-g}\,\phi\left(\frac{1}{2}D_{ab}+\frac{R+2\square}{3}g_{ab}\right)F_{2}\left(\square\right)\frac{1}{2}\left(\square-\frac{R}{3}\right)h^{\bot ab}\,.
\end{equation}
The metric $g_{ab}$ can be commuted through to annihilate $h^{\bot ab}$,
so the only potentially problematic term is of the form
\[
\nabla_{a}\nabla_{b}F_{2}\left(\square\right)\left(\square-\frac{R}{3}\right)h^{\bot ab}.
\]
By expanding $F_{2}\left(\square\right)$ in its power series $F_{2}=\overset{\infty}{\underset{n=0}{\sum}}f_{2n}\square^{n}$
and then using the commutation relation Eq.~(\ref{eq:commutation 4})
iteratively, we can commute through $\nabla_{b}$ all the way till
it annihilates $h^{\bot ab}$. Hence, we have shown that the physical
fields decouple nicely and we can turn now to the evaluation of the
propagators. To start with the scalar mode, we use the expressions
\begin{align}
\delta R_{ab}\left(\phi\right) & =\left(\frac{1}{2}D_{ab}+\frac{R+2\square}{3}g_{ab}\right)\phi\,,\nonumber \\
\delta R\left(\phi\right) & =\left(2\square+R\right)\phi\,,\nonumber \\
\widetilde{\delta}_{0}\left(\phi\right) & =-\frac{1}{4}\phi\left(2\square+R\right)\phi\,,
\end{align}
derived above to write
\begin{eqnarray}
 & S_{qua}\left(\phi\right)= & \int d^{3}x\,\sqrt{-g}\mbox{\ensuremath{\Biggr[}}-\left(\frac{1}{8}+\frac{1}{2}\widehat{f}_{10}R\right)\phi\left(2\square+R\right)\phi+\phi\left(2\square+R\right)F_{1}\left(\square\right)\left(2\square+R\right)\phi\nonumber \\
 &  & +\phi\left(\frac{1}{2}D_{ab}+\frac{R+2\square}{3}g_{ab}\right)F_{2}\left(\square\right)\left(\frac{1}{2}D^{ab}+\frac{R+2\square}{3}g^{ab}\right)\phi\mbox{\ensuremath{\Biggr]}}.
\end{eqnarray}
The last term still allows for some simplification. After expanding
$F_{2}\left(\square\right)$ the commutation relation Eq.~(\ref{eq:commutation 1})
can be used to obtain: %
\begin{equation}
\int d^{3}x\,\sqrt{-g}\left[\phi D_{ab}F_{2}\left(\square\right)D^{ab}\phi\right]=\int d^{3}x\,\sqrt{-g}\left[\phi F_{2}\left(\square+R\right)\left(\frac{2\square+R}{3}\right)\square\phi\right].
\end{equation}
Now, putting everything together and using the tracelessness of $D_{ab}$,
see Eq.(\ref{Dab}), yields the final result
\begin{align}
S_{qua}\left(\phi\right)= & \int d^{3}x\,\sqrt{-g}\,\phi\mbox{\ensuremath{\Biggr[}}-\left(\frac{1}{8}+\frac{1}{2}\widehat{f}_{10}R\right)+F_{1}\left(\square\right)\left(2\square+R\right)+\nonumber \\
 & \frac{1}{3}F_{2}\left(\square\right)\left(2\square+R\right)+\frac{1}{12}F_{2}\left(\square+R\right)\square\mbox{\ensuremath{\Biggr]}}\left(2\square+R\right)\phi.
\end{align}
The tensor mode is a lot simpler to handle because of its transverse-traceless
property. Using the expressions Eq. (\ref{eq:deltaRab hab}, \ref{eq:delta0 hab})
the final action results in
\begin{equation}
S_{qua}\left(h^{\bot}\right)=\int d^{3}x\,\sqrt{-g}\,\frac{1}{8}h_{ab}^{\bot}\left[\left(\square-\frac{R}{3}\right)\left(1+4\widehat{f}_{10}R\right)+2F_{2}\left(\square\right)\left(\square-\frac{R}{3}\right)^{2}\right]h^{\bot ab}.
\end{equation}
The tensor and scalar part of the propagators can now be given by:
\begin{equation}
\Pi\left(\phi\right)=\frac{P_{s}^{0}}{\left[1+4\widehat{f}_{10}R-16F_{1}\left(\square\right)\left(\square+\frac{R}{2}\right)-\frac{16}{3}F_{2}\left(\square\right)\left(\square+\frac{R}{2}\right)-\frac{2}{3}F_{2}\left(\square+R\right)\square\right]\left(\square+\frac{R}{2}\right)}\,,\label{eq:phi propagator}
\end{equation}
and
\begin{equation}
\Pi\left(h^{\bot}\right)=-\frac{P_{s}^{2}}{\left[1+4\widehat{f}_{10}R+2F_{2}\left(\square\right)\left(\square-\frac{R}{3}\right)\right]\left(\square-\frac{R}{3}\right)}\,.\label{eq:h propagator}
\end{equation}
An important issue here is the normalization of the propagators: Since
we want to take the Minkowski limit $R\rightarrow0$ in the next section,
the propagators have to be normalized correctly. From Eq. (\ref{eq:propagator})
and Eq. (\ref{eq:abcdf relations}) we see that the first constant
term in the denominater of $\Pi\left(h^{\bot}\right)$ should be 1,
hence we chose that as our normalization condition and removed a factor
of $\frac{1}{8}$ from both propagators. For the scalar part\textbf{
}we had to add an additional factor of $\frac{1}{2}$ which is contained
in the spin projection operator.


\subsection{Discussions, comparisons and IR limits}

We turn now to the interpretation of the results obtained in Eqs.~(\ref{eq:phi propagator},\ref{eq:h propagator}).
As a nice cross-check we can take the limit $R\rightarrow0$, which
should of course reproduce the propagator in Minkowski space. We get
\begin{equation}
\underset{\Lambda=0}{\Pi}\left(h^{\bot}\right)=-\frac{1}{\square\left(1+2F_{2}\left(\square\right)\square\right)}=-\frac{1}{a\left(\square\right)\square}\,,
\end{equation}
by using the relations Eq.~(\ref{eq:abcdf relations}) which is the
desired result. For the scalar part\textbf{ }we have to add an additional
factor of $\frac{1}{2}$ which is contained in the spin projection
operator but after that we get
\begin{equation}
\underset{\Lambda=0}{\Pi}\left(\phi\right)=-\frac{1}{\square\left(-1+16F_{1}\left(\square\right)\square+6F_{2}\left(\square\right)\square\right)}=-\frac{1}{\left(a\left(\square\right)-2c\left(\square\right)\right)\square}.
\end{equation}
as expected.

Another instructive limit is to take is $F_{i}\left(\square\right)\rightarrow0$,
in Eqs.~(\ref{eq:phi propagator},\ref{eq:h propagator}), i.e. turning
off the infinite derivative terms. This would leave the graviton off-shell
propagator in the (A)ds background: %
\begin{equation}
\Pi=\frac{P_{s}^{2}}{-\square+\frac{R}{3}}-\frac{P_{s}^{0}}{-\square-\frac{R}{2}}\,,\label{eq:limit propagator}
\end{equation}
where we see that the graviton acquires a non-vanishing mass due to
the spacetime curvature.

The last question remains is what is the most natural choice for the
form factors $F_{i}\left(\square\right)$? In flat space we required
the propagator to be proportional to the GR-propagator and this should
also be our goal here. By comparing Eqs.~(\ref{eq:phi propagator},
\ref{eq:h propagator}) with Eq.~(\ref{eq:limit propagator}), we
obtain the necessary constraint on the form factors:
\begin{equation}
F_{1}\left(\square\right)=-\frac{1}{8}F_{2}\left(\square\right)\frac{\left(\square-\frac{R}{3}\right)}{\left(\square+\frac{R}{2}\right)}-\frac{1}{24}F_{2}\left(\square+R\right)\frac{\square}{\left(\square+\frac{R}{2}\right)}-\frac{1}{3}F_{2}\left(\square\right)\,.\label{eq:F1 F2 relation}
\end{equation}
In the Minkowski limit, when $R=0$, we obtain exactly $2F_{1}(\Box)+F_{2}(\Box)=0$,
see Eq.~(\ref{eq:constraint}).

Now, back into the (A)ds, the function of proportionality which we
shall call $a\left(\square\right)$ in accordance with the chapter
3 (the treatment in the Minkowski space) is given by %
\begin{equation}
a\left(\square\right)=1+4\widehat{f}_{10}R+2F_{2}\left(\square\right)\left(\square-\frac{R}{3}\right).\label{eq:a F relation}
\end{equation}
For not introducing any new zeros in the propagator, $a\left(\square\right)$
has to be an exponential of an entire function, i.e. $Ce^{\gamma\left(\square\right)}$
where $\gamma\left(\square\right)$ is an entire function and $C\neq0$
a constant. However, a simple choice of $a\left(\square\right)=e^{-\frac{\square}{M_{s}^{2}}}$
is not viable anymore, because $F_{1}\left(\square\right)$ and $F_{2}\left(\square\right)$
will not be analytic then: If we solve Eq. (\ref{eq:a F relation})
for $F_{2}\left(\square\right)$ we get
\begin{equation}
F_{2}\left(\square\right)=\frac{Ce^{\gamma\left(\square\right)}-1-4\widehat{f}_{10}R}{2\left(\square-\frac{R}{3}\right)}\,.
\end{equation}
If we expand the exponential we see that we must have
\begin{equation}
C=1+4\widehat{f}_{10}R\,,
\end{equation}
otherwise we produce a term proportional to $\frac{1}{\left(\square-\frac{R}{3}\right)}$.
Moreover, $\gamma\left(\square\right)$ has to contain a factor $\left(\square-\frac{R}{3}\right)$
to cancel the denominator, so we arrive at
\begin{equation}
a\left(\square\right)=\left(1+4\widehat{f}_{10}R\right)e^{\left(\square-\frac{R}{3}\right)\tau\left(\square\right)}
\end{equation}
with some entire function $\tau\left(\square\right)$. Now, solving
Eq. (\ref{eq:F1 F2 relation}) for $F_{1}\left(\square\right)$ yields
\begin{align}
F_{1}\left(\square\right)= & -\frac{1}{16}\frac{\left(1+4\widehat{f}_{10}R\right)\left(e^{\left(\square-\frac{R}{3}\right)\tau\left(\square\right)}-1\right)}{\square+\frac{R}{2}}-\frac{1}{48}\frac{\left(1+4\widehat{f}_{10}R\right)\left(e^{\left(\square+\frac{2R}{3}\right)\tau\left(\square+R\right)}-1\right)\square}{\left(\square+\frac{R}{2}\right)\left(\square+\frac{2R}{3}\right)}\nonumber \\
 & -\frac{1}{6}\frac{\left(1+4\widehat{f}_{10}R\right)\left(e^{\left(\square-\frac{R}{3}\right)\tau\left(\square\right)}-1\right)}{\square-\frac{R}{3}}
\end{align}
and again analyticity demands that we cancel the denominators. We
see that $\tau\left(\square\right)$ has to contain the factors $\square+\frac{R}{2}$
and $\square-\frac{R}{2}$, hence the simplest choice for $a\left(\square\right)$
is
\begin{equation}
a\left(\square\right)=\left(1+4\widehat{f}_{10}R\right)e^{-\frac{\left(\square+\frac{R}{2}\right)\left(\square-\frac{R}{2}\right)\left(\square-\frac{R}{3}\right)}{M_{s}^{6}}}\,,
\end{equation}
then the $F_{i}\left(\square\right)$ will be perfectly analytic:
\begin{eqnarray}
F_{1}\left(\square\right) & = & -\left(1+4\widehat{f}_{10}R\right)\mbox{\ensuremath{\Biggl(}}\frac{e^{-\frac{\left(\square+\frac{R}{2}\right)\left(\square-\frac{R}{2}\right)\left(\square-\frac{R}{3}\right)}{M_{s}^{6}}}-1}{16\left(\square+\frac{R}{2}\right)}+\nonumber \\
 &  & \frac{\left(e^{-\frac{\left(\square+\frac{3R}{2}\right)\left(\square+\frac{R}{2}\right)\left(\square+\frac{2R}{3}\right)}{M_{s}^{6}}}-1\right)\square}{48\left(\square+\frac{2R}{3}\right)\left(\square+\frac{R}{2}\right)}+\frac{e^{-\frac{\left(\square+\frac{R}{2}\right)\left(\square-\frac{R}{2}\right)\left(\square-\frac{R}{3}\right)}{M_{s}^{6}}}-1}{6\left(\square-\frac{R}{3}\right)}\mbox{\ensuremath{\Biggr)}}\,,\label{eq:1st}
\end{eqnarray}
and
\begin{equation}
F_{2}\left(\square\right)=\frac{\left(1+4\widehat{f}_{10}R\right)\left(e^{-\frac{\left(\square+\frac{R}{2}\right)\left(\square-\frac{R}{2}\right)\left(\square-\frac{R}{3}\right)}{M_{s}^{6}}}-1\right)}{2\left(\square-\frac{R}{3}\right)}\,.\label{2nd}
\end{equation}
There is one last point to consider: Note that we have treated $\widehat{f}_{10}$
as an independent variable so far, however, it is supposed to be the
zeroth order coefficient of $\widehat{F}_{1}\left(\square\right)$.
With the help of Eqs.~(\ref{redefine}, \ref{eq:1st}), we obtain:
\begin{equation}
\widehat{F}_{1}\left(\square\right)=-\left(1+4\widehat{f}_{10}R\right)\left(\frac{e^{-\frac{\left(\square+\frac{R}{2}\right)\left(\square-\frac{R}{2}\right)\left(\square-\frac{R}{3}\right)}{M_{s}^{6}}}-1}{16\left(\square+\frac{R}{2}\right)}+\frac{\left(e^{-\frac{\left(\square+\frac{3R}{2}\right)\left(\square+\frac{R}{2}\right)\left(\square+\frac{2R}{3}\right)}{M_{s}^{6}}}-1\right)\square}{48\left(\square+\frac{2R}{3}\right)\left(\square+\frac{R}{2}\right)}\right)\,.
\end{equation}
We can now extract the zeroth order term from above, and obtain
\begin{equation}
\widehat{f}_{10}=-\left(1+4\widehat{f}_{10}R\right)\frac{e^{-\frac{R^{3}}{12M_{s}^{6}}}-1}{8R}\,,
\end{equation}
with the solution
\begin{equation}
\widehat{f}_{10}=\frac{1}{4R}\frac{1-e^{-\frac{R^{3}}{12M_{s}^{6}}}}{1+e^{-\frac{R^{3}}{12M_{s}^{6}}}}.
\end{equation}
We should point out that the form factors depend explicitly on the
background curvature. If one takes the limit $R\rightarrow0$ in the
above expressions, we will reduce $F_{1}(\Box)$ and $F_{2}(\Box)$
to
\begin{equation}
F_{1}\left(\square\right)=-\frac{e^{-{\square^{3}}/{M_{s}^{6}}}-1}{4\square},\,\,\,\,\,\,\quad\quad F_{2}\left(\square\right)=\frac{e^{-{\square^{3}}/{M_{s}^{6}}}-1}{2\square}
\end{equation}
which defers from the flat space case in the sense that we have different
powers of $\frac{\square}{M_{s}^{2}}$, see Eq.~(\ref{eq:F_in flat space}).
If we want to have a smooth Minkowski limit we have to replace $a\left(\square\right)$
in Eq. (\ref{eq:a box flat}) by
\begin{equation}
a'\left(\square\right)=e^{-{\square^{3}}/{M_{s}^{6}}}\,.
\end{equation}
By the above choice, the desirable properties of the theory like tree
level unitarity will remain unchanged by this modification.


\section{Conclusion}

This paper provides the ghost free conditions for parity invariant
and torsion free AIDG in $3d$. At first we determined the full equations
of motion and deduced the linearized limit in flat space in complete
analogy to the $4d$ case without introducing any new degrees of freedom.
As expected, we also found exact maximally symmetric solutions which
can serve as background solutions for linearization. With considerable
algebraic effort, it was possible to construct a well defined linearized
theory around those (A)dS-backgrounds. We also considered New Massive
Gravity as a low-energy limit instead of GR, and succeeded in constructing
an AIDG action around new massive gravity around the Minkowski background.

The main highlights of the paper are following. First of all we have
shown that the vacuum solution of AIDG in $3$ dimensions respects
the BTZ blackhole solution in AdS. However, adding a point source
generates a non-trivial, non-singular solution. The solution so far
has been obtained only around the Minkowski background. Second important
result is that we have derived two main equations in this paper containing
the scalar and the graviton propagators for AIDG action in (A)dS in
$3$ dimensions, see Eqs.(\ref{eq:phi propagator},\ref{eq:h propagator}).
These have been obtained by perturbing the action up to quadratic
in metric potential, i.e. ${\cal O}(h^{2})$ around (A)dS background
in $3$ dimensions. We have discussed various consistency checks,
such as our results of the propagators match the expectations around
the Minkowski background. We have also verified that the propagator
reduces to that of Einstein gravity in $3$ dimensions around the
(A)dS background when we take the appropriate limit $\Box/M_{s}^{2}\rightarrow0$,
or $F_{i}(\Box)\rightarrow0$. We have also provided an example of
the analytic form factors $F_{1}(\Box)$ and $F_{2}(\Box)$ around
(A)dS backgrounds.

There are still some open questions remain. First, we have not proven
that the maximally symmetric spacetimes are really the only vacuum
solutions. If that is the case, it would be natural to assume that
AIDG in the vacuum, as GR, is a topological field theory. Since it
does not seem to be a Chern-Simons-theory (at least there is no natural
connection) yet, it is an interesting open problem to classify it
as some other topological field theory. Furthermore, one could add
a boundary and try to find the dual conformal field theory, if it
exists, to provide a new realization of the holographic principle.
All in all, AIDG in three dimensions has shown many interesting features
which make it worth studying these aspects further.\\

\textbf{Acknowledgements} We would like to thank Raphaela Wutte and
Masahide Yamaguchi for helpful discussions. AM's research is funded
by the Netherlands Organisation for Scientific Research (NWO) grant
number 680-91-119. AM would like to thank JSPS for their hospitality
and Masahide Yamaguchi from the Tokyo Institute of Technology for
hosting me where part of the work has been carried out.

\section{Appendix}

\subsection{Inverting the field equations}

We can write the linearized field equations in the form $\left(\Pi^{-1}\right)_{ab}^{\,\,\,\,\,\, cd}h_{cd}=\kappa\tau_{ab}$
with the linear operator

\begin{align}
\left(\Pi^{-1}\right)_{ab}^{\,\,\,\,\,\, cd} & =k^{2}a\left(-k^{2}\right)\delta_{a}^{(c}\delta_{b}^{d)}+2b\left(-k^{2}\right)k^{(c}k_{(a}\eta_{b)}^{d)}+\nonumber \\
 & c\left(-k^{2}\right)\left(\eta_{ab}k^{c}k^{d}+k_{a}k_{b}\eta^{cd}\right)+k^{2}d\left(-k^{2}\right)\eta_{ab}\eta^{cd}+\frac{f\left(-k^{2}\right)}{k^{2}}k_{a}k_{b}k^{c}k^{d}.\label{eq:eom-operator}
\end{align}
One can significally simplify $\Pi^{-1}$ by using invariance properties:
Eq.~(\ref{eq:eom-operator}) is constructed solely out of $\eta_{ab}$
and $k^{a}$, hence the little group of $k^{a}$ commutes with it.
If we take $k^{a}$ to be time-like, then the little group is SO(2)~
\footnote{Later it will turn out that $k^{a}$ is actually light-like, however,
SO(2) is more useful to decompose the eoms than the little group of
a light-like vector, ISO(1).%
}. By Schur's lemma, every operator that commutes with all elements
of a group in one of its irreducible representations, has to be proportional
to the identity operator. The symmetric-tensor-representation of SO(2)
is decomposable into four irreducible representations: one with spin
two (2 degrees of freedom), one with spin one (2 dof) and two scalars
(1 dof each).%
\footnote{In three dimensions, the notion of \emph{spin }should be regarded
with care: Usually, spin refers to representations of SO(3) which
are important for massive particles in four dimensions. The representations
of SO(2) we use here are the same as for massless particles in 4d
which we classify according to their \emph{helicity.} %
} It is now useful to define the so-called \emph{spin projection operators
}which\emph{ }project on these four subspaces~\cite{VanNieuwenhuizen:1973fi,Biswas:2013kla}:\emph{
\begin{align}
 & P_{s}^{2}=\frac{1}{2}\left(\theta_{ac}\theta_{bd}+\theta_{ad}\theta_{bc}\right)-\frac{1}{2}\theta_{ab}\theta_{cd}\,,\nonumber \\
 & P_{w}^{1}=\frac{1}{2}\left(\theta_{ac}\omega_{bd}+\theta_{ad}\omega_{bc}+\theta_{bc}\omega_{ad}+\theta_{bd}\omega_{ac}\right)\,,\nonumber \\
 & P_{s}^{0}=\frac{1}{2}\theta_{ab}\theta_{cd}\,,\nonumber \\
 & P_{w}^{0}=\omega_{ab}\omega_{cd}\,,
\end{align}
}with
\begin{equation}
\theta_{ab}=\eta_{ab}-\frac{k_{a}k_{b}}{k^{2}}\,\,\,\mbox{and}\,\,\,\omega_{ab}=\frac{k_{a}k_{b}}{k^{2}}.
\end{equation}
It is easy to verify that they are all orthogonal and satisfy: %
\begin{equation}
P_{s}^{2}+P_{w}^{1}+P_{s}^{0}+P_{w}^{0}=1\,\,\,\mbox{and}\,\,\, P_{a}^{i}P_{b}^{j}=\delta^{ij}\delta_{ab}.
\end{equation}
$P_{s}^{2}$ corresponds to the transverse and traceless degrees of
freedom, $P_{w}^{1}$ to the longitudinal and traceless ones, $P_{s}^{0}$
represents the transverse trace part and $P_{w}^{0}$ the scalar which
is neither transverse nor traceless. The operator Eq.~(\ref{eq:eom-operator})
has to be proportional to unity in each subspace and must not mix
subspaces of different spin, however, it could potentially mix $P_{s}^{0}$
with $P_{w}^{0}$. Fortunately, that does not happen in our case and
we can write
\begin{equation}
\left(\Pi^{-1}\right)_{ab}^{\,\,\,\,\,\, cd}=AP_{s}^{2}+BP_{w}^{1}+CP_{s}^{0}+DP_{w}^{0}\,.
\end{equation}
Determining the coefficients is straight-forward, one gets
\begin{equation}
\Pi^{-1}=k^{2}aP_{s}^{2}+k^{2}\left(a-2c\right)P_{s}^{0},
\end{equation}
so the longitudinal parts $P_{w}^{1}$ and $P_{w}^{0}$ simply dropped
out. That implies that the energy-momentum-tensor $\tau_{ab}$ also
must not have any longitudinal modes, it has to be conserved: $k^{a}\tau_{ab}=0$.
So if we just remove the longitudinal degrees of freedom from our
solution space, the equations of motion can be inverted and the resulting
propagator is
\begin{equation}
\Pi_{AIDG}=\frac{P_{s}^{2}}{ak^{2}}+\frac{P_{s}^{0}}{\left(a-2c\right)k^{2}}\,.
\end{equation}
If we want $\Pi_{AIDG}$ to be proportional to the $\Pi_{GR}$ such
that no additional particles are introduced we demand $a=c$ (or equivalently
$f=0$) to obtain
\begin{equation}
\Pi_{AIDG}=\frac{1}{k^{2}a\left(-k^{2}\right)}\left(P_{s}^{2}-P_{s}^{0}\right)\,.
\end{equation}


\subsection{Degrees of freedom in Einstein gravity }

It can be shown easily that Eq.~(\ref{eq: lin eom in mom space})
with $\tau_{ab}=0$ imply $k^{2}=0$. After removing the $k^{2}$-terms
we see that both $k^{a}h_{ab}$ and $h$ have to be zero, i. e. $h_{ab}$
has to be transverse and traceless. We can expand $h_{ab}$ in a light-like
basis using $k_{a}$ and additionally a second light-like vector $l_{a}$
and an orthogonal space-like vector $e_{a}$ as basis vectors. The
Minkowski metric then takes the form %
\begin{equation}
\eta_{ab}=\begin{pmatrix}0 & -1 & 0\\
-1 & 0 & 0\\
0 & 0 & 1
\end{pmatrix}\,,
\end{equation}
and $h_{ab}$ can be written as
\begin{equation}
h_{ab}=\alpha\left(k\right)k_{a}k_{b}+2\beta\left(k\right)k_{(a}l_{b)}+\gamma\left(k\right)l_{a}l_{b}+2\delta\left(k\right)k_{(a}e_{b)}+2\phi\left(k\right)l_{(a}e_{b)}+\lambda\left(k\right)e_{a}e_{b}\,,
\end{equation}
with $k$-dependent coefficients. Transverse traceless now means that
$\beta=\gamma=\phi=\lambda=0.$ The remaining coeffiecients are gauge
degrees of freedom and can be removed by a gauge transformation
\begin{equation}
h_{ab}\rightarrow h_{ab}+k_{a}v_{b}+v_{a}k_{b}\,,\label{eq:gauge trafo}
\end{equation}
with some vector $v_{a}$.


\subsection{Perturbations}

Here we summarize the expressions for the perturbations up to second
order of all relevant geometrical quantities; background quantities
are indicated with a bar. The basic definition is
\begin{equation}
g_{ab}=\overline{g}_{ab}+h_{ab},
\end{equation}
raising and lowering indices is always done using $\overline{g}_{ab}.$
It follows
\begin{equation}
g^{ab}\approx\overline{g}^{ab}-h^{ab}+h^{ac}h_{c}^{b},\,\,\,\,\,\,\,\,\,\sqrt{-g}\approx\sqrt{-\overline{g}}\left(1+\frac{h}{2}+\frac{h^{2}}{8}-\frac{h_{ab}h^{ab}}{4}\right)\,,
\end{equation}
\begin{equation}
\Gamma_{bc}^{a}\approx\overline{\Gamma}_{bc}^{a}+\delta\Gamma_{bc}^{a},\,\,\,\,\,\,\,\,\,\delta\Gamma_{bc}^{a}=\frac{1}{2}\left(\overline{\nabla}_{b}h_{c}^{a}+\overline{\nabla}_{c}h_{b}^{a}-\overline{\nabla}^{a}h_{bc}\right),
\end{equation}
\begin{equation}
\delta R_{\,\,\,\,\, cd}^{ab}=\frac{\overline{R}}{12}\left(\delta_{d}^{a}h_{c}^{b}-\delta_{c}^{a}h_{d}^{b}+\delta_{c}^{b}h_{d}^{a}-\delta_{d}^{b}h_{c}^{a}\right)+\frac{1}{2}\left(\bar{\nabla}_{c}\bar{\nabla}^{b}h_{d}^{a}-\overline{\nabla}_{c}\overline{\nabla}^{a}h_{d}^{b}-\overline{\nabla}_{d}\overline{\nabla}^{b}h_{c}^{a}+\overline{\nabla}_{d}\overline{\nabla}^{a}h_{c}^{b}\right)\,,\label{eq: Riemann pert}
\end{equation}
\begin{equation}
R_{ab}\approx\overline{R}_{ab}+\delta R_{ab}+\delta^{2}R_{ab},\,\,\,\,\,\,\,\,\,\delta R_{ab}=\overline{\nabla}_{c}\delta\Gamma_{ab}^{c}-\overline{\nabla}_{b}\delta\Gamma_{ac}^{c},\,\,\,\,\,\,\,\delta^{2}R_{ab}=\delta\Gamma_{dc}^{c}\delta\Gamma_{ab}^{d}-\delta\Gamma_{db}^{c}\delta\Gamma_{ca}^{d}\,,\label{eq:Ricci pert}
\end{equation}
\begin{equation}
R\approx\overline{R}+\delta R,\,\,\,\,\,\,\,\,\,\delta R=-h^{ab}\overline{R}_{ab}+\overline{g}^{ab}\left(\overline{\nabla}_{c}\delta\Gamma_{ab}^{c}-\overline{\nabla}_{b}\delta\Gamma_{ac}^{c}\right)\,.
\end{equation}
The order of expansion in $h_{ab}$ is either first or second, depending
on what we need to vary the action.


\subsection{Quadratic action for Einstein gravity}

Obtaining the second variation of $S_{EH}$ in a curved backgroound
is a straight forward, but laborious task. We start by expanding every
quantity up to second order:
\begin{equation}
S_{EH}=\int d^{3}x\,\sqrt{-g}\left(1+\frac{h}{2}+\frac{h^{2}}{8}-\frac{h_{ab}h^{ab}}{4}\right)\left[\frac{1}{2}\left(R_{ab}+\delta R_{ab}+\delta^{2}R_{ab}\right)\left(g^{ab}-h^{ab}+h^{ac}h_{c}^{b}\right)-\Lambda\right]\,.
\end{equation}
Collecting all the quadratic terms yields
\begin{eqnarray}
\delta^{2}S_{EH} & = & \int d^{3}x\,\sqrt{-g}\mbox{\ensuremath{\Biggr[}}\frac{1}{2}R_{ab}h^{ac}h_{c}^{b}-\frac{1}{2}\delta R_{ab}h^{ab}+\frac{1}{2}\delta^{2}R_{ab}g^{ab}\nonumber \\
 &  & -\frac{h}{4}R_{ab}h^{ab}+\frac{h}{4}\delta R_{ab}g^{ab}+\left(\frac{h^{2}}{8}-\frac{h_{ab}h^{ab}}{4}\right)\left(\frac{R}{2}-\Lambda\right)\Biggl]\,.
\end{eqnarray}
Plugging in the perturbations results in
\begin{eqnarray}
\delta^{2}S_{EH} & = & \int d^{3}x\,\sqrt{-g}\mbox{\ensuremath{\Biggr[}}\frac{1}{6}h^{ab}h_{ab}-\frac{1}{4}\left(\nabla_{c}\nabla_{a}h_{b}^{c}-\nabla_{c}\nabla_{b}h_{a}^{c}-\square h_{ab}+\nabla_{a}\nabla_{b}h\right)h^{ab}\nonumber \\
 &  & +\frac{1}{4}\nabla_{b}h\left(\nabla_{a}h^{ab}-\frac{1}{2}\nabla^{b}h\right)-\frac{1}{8}\left(\nabla_{b}h_{c}^{a}+\nabla_{c}h_{b}^{a}-\nabla^{a}h_{bc}\right)\left(\nabla_{a}h^{bc}+\nabla^{c}h_{a}^{b}-\nabla^{b}h_{a}^{c}\right)\nonumber \\
 &  & -\frac{1}{12}h^{2}R+\frac{h}{4}\left(\nabla_{a}\nabla_{b}h^{ab}-\square h\right)+\left(\frac{h^{2}}{8}-\frac{h_{ab}h^{ab}}{4}\right)\left(\frac{R}{2}-\Lambda\right)\Biggl]\nonumber \\
 & = & \int d^{3}x\,\sqrt{-g}\Biggr[\frac{1}{8}h^{ab}\square h_{ab}-\frac{1}{8}h\square h+\frac{1}{4}h\nabla_{a}\nabla_{b}h^{ab}\nonumber \\
 &  & -\frac{1}{4}h^{ab}\nabla_{c}\nabla_{a}h_{b}^{c}-\frac{R}{48}h^{2}+\frac{R}{24}h_{ab}h^{ab}-\Lambda\left(\frac{h^{2}}{8}-\frac{h_{ab}h^{ab}}{4}\right)\Biggl]\,.
\end{eqnarray}
The fourth term can be further modified
\begin{eqnarray}
-\frac{1}{4}h^{ab}\nabla_{c}\nabla_{a}h_{b}^{c} & = & \frac{1}{4}\nabla_{a}h^{ab}\nabla_{c}h_{b}^{c}-\frac{1}{4}h^{ab}R_{\,\,\, bac}^{d}h_{d}^{c}-\frac{1}{4}h^{ab}R_{\,\,\, dca}^{c}h_{b}^{d}\nonumber \\
 & = & \frac{1}{4}\nabla_{a}h^{ab}\nabla_{c}h_{b}^{c}-\frac{R}{24}h_{ab}h^{ab}+\frac{R}{24}h^{2}-\frac{R}{12}h_{ab}h^{ab}\,,
\end{eqnarray}
such that the final result becomes
\begin{eqnarray}
\delta^{2}S_{EH} & = & \int d^{3}x\,\sqrt{-g}\Biggr[\frac{1}{8}h^{ab}\square h_{ab}-\frac{1}{8}h\square h+\frac{1}{4}h\nabla_{a}\nabla_{b}h^{ab}\nonumber \\
 &  & +\frac{1}{4}\nabla_{a}h^{ab}\nabla_{c}h_{b}^{c}+\frac{R}{48}h^{2}-\frac{R}{12}h_{ab}h^{ab}-\Lambda\left(\frac{h^{2}}{8}-\frac{h_{ab}h^{ab}}{4}\right)\Biggl].
\end{eqnarray}

\subsection{Commutation relations}

We list here some useful commutation relations for differential operators
which hold on maximally symmetric backgrounds:
\begin{equation}
\nabla_{a}\square t^{a}=\left(\square+\frac{R}{3}\right)\nabla_{a}t^{a}\,,\label{eq:commutation 2}
\end{equation}
for a generic vector $t^{a}$,
\begin{equation}
\nabla_{a}\square t^{ab}=\left(\square+\frac{2R}{3}\right)\nabla_{a}t^{ab}-\frac{R}{3}\nabla^{b}t_{a}^{a}\,,\label{eq:commutation 4}
\end{equation}
for symmetric tensors $t^{ab}$, and
\begin{equation}
\nabla_{a}\nabla^{b}A^{a}=\frac{R}{3}A^{b}\,,\label{eq:commutation 3}
\end{equation}
for transverse vectors $A^{a}$. In general,
\begin{equation}
\nabla_{a}\nabla_{b}\square^{n}D^{ab}\phi=\left(\square+R\right)^{n}\left(\frac{2\square+R}{3}\right)\square\phi\,,\label{eq:commutation 1}
\end{equation}
holds for the operator $D_{ab}$ defined in section 5. All of those
relations can be derived by straight forward Riemann tensor substitution.


\end{document}